\documentclass[11pt,twoside]{article}
\usepackage{asp2010}

\resetcounters

\begin{document}

\title{The long term dynamics of the solar radiative zone associated \\to new results from SoHO and young solar analogs}
\author{Sylvaine~Turck-Chi\`eze$^1$, S\' ebastien~Couvidat$^2$, Antonio Eff-Darwich$^3$, \\Vincent Duez$^4$, Rafael A. Garc\`ia$^1$, St\' ephane Mathis$^1$, Savita Mathur$^5$, \\Laurent Piau$^6$, David Salabert$^7$
\affil{$^1$CEA/DSM/IRFU/SAp-AIM, CE Saclay, 
Orme des  Merisiers,  91191 Gif-sur-Yvette, France}
\affil{$^2$W.W. Hansen. E. P. L., Stanford University, Stanford, CA 94305, USA}
\affil{$^3$IAC, La Laguna, T\' en\' erife, Spain}
\affil{$^4$Argelander Institut fur Astronomie, Universitat Bonn, Germany}
\affil{$^5$High Altitude Observatory, NCAR, P.O. Box 3000, Boulder, CO 80307, USA}
\affil{$^6$LATMOS, 78 St Quentin en Yvelines, France}
\affil{$^7$Universit\'e de Nice Sophia-Antipolis, CNRS, Observatoire de la C\^ote d'Azur, BP 4229, 06304 Nice Cedex 4, France}}
\begin{abstract}
The Standard Solar Model  (SSM) is no more sufficient to interpret all the observations of the radiative zone obtained with the SoHO satellite. We recall  our present knowledge of this internal region and compare the recent results to models beyond the SSM assumptions. Then we discuss the missing processes and quantify some of them in using young analog observations to build a more realistic view of our star. This progress will be useful for solar-like stars observed by COROT and KEPLER.
\end{abstract}
\vspace{-1cm}
\section{A dynamical view of the radiative zone of Sun and solar-like stars}
The SoHO satellite (Domingo 1995)  continues to measure the solar acoustic mode characteristics and has also detected the first gravity mode frequencies \citep{Turck2001, Turck2004,Garcia2007,Garcia2011}. From these data, we deduce a sound speed profile and  a rotation profile in the whole radiative zone. 

On the other side, neutrino detections put other constraints on the solar core  in total agreement with the helioseismic results (Turck-Chi\`eze and Couvidat 2011).  Moreover the study of gravitational moments shows also that the dynamical processes cannot be forgotten and that even for this surface effect, the knowledge of the dynamics of the radiative zone is crucial (Duez, Turck-Chi\`eze, Mathis 2011).

Stellar internal radiative zones depend on  the long term  history of stars, their energy transfer, rotation history and  activity of their early stages. The Sun is the only star for which the internal radiative zone is known with  such details but hopefully, thousand analogous stars begin to be studied by asteroseismology.  Young analogs  are also observed in UV and EUV with their amount of mass loss, some magnetic configuration reveals the topology of their magnetic field. So these complement observations invite us to create the link between young stars and solar-like stars of several Gyrs.

\section{The knowledge of the internal radiative zone}
\vspace{-0.3cm}
\subsection{Solar sound speed, density and rotation profiles }
The knowledge of the solar radiative zone has been largely improved thanks to the GOLF \citep{Gabriel} and MDI \citep{Scherrer} instruments aboard SoHO. 
The detection of most of the acoustic modes, including the radial modes $\ell$ = 0 of low orders that are not polluted by any surface solar cycle effect, has allowed a determination of the sound speed down to 0.06 R$_\odot$ without ambiguity \citep{Turck2001}  in integrating 5 years of SOHO/ GOLF and MDI observations. The inversion of these data leads to an extremely precise vertical error bar but a non negligible horizontal error bar (see also Figure 2 left). This profile has been confirmed after 30 years of  measurements on ground by the  BiSON network for the low degree modes \citep{Basu2009}.  The two results are strictly the same in the radiative zone. 

 \begin{figure*}
\includegraphics[width=13pc] {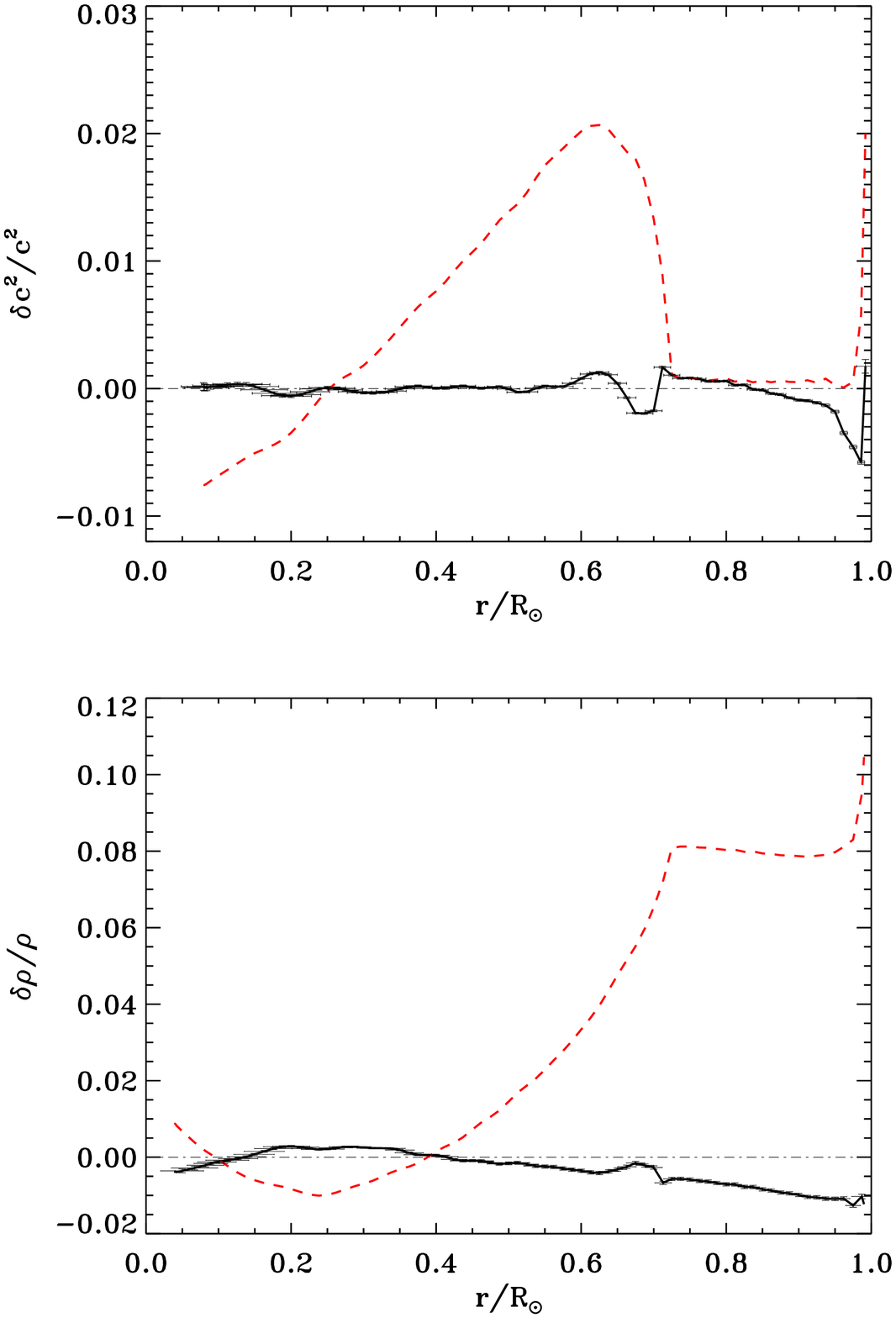}
\vspace{-1pc}
\includegraphics[width=20pc] {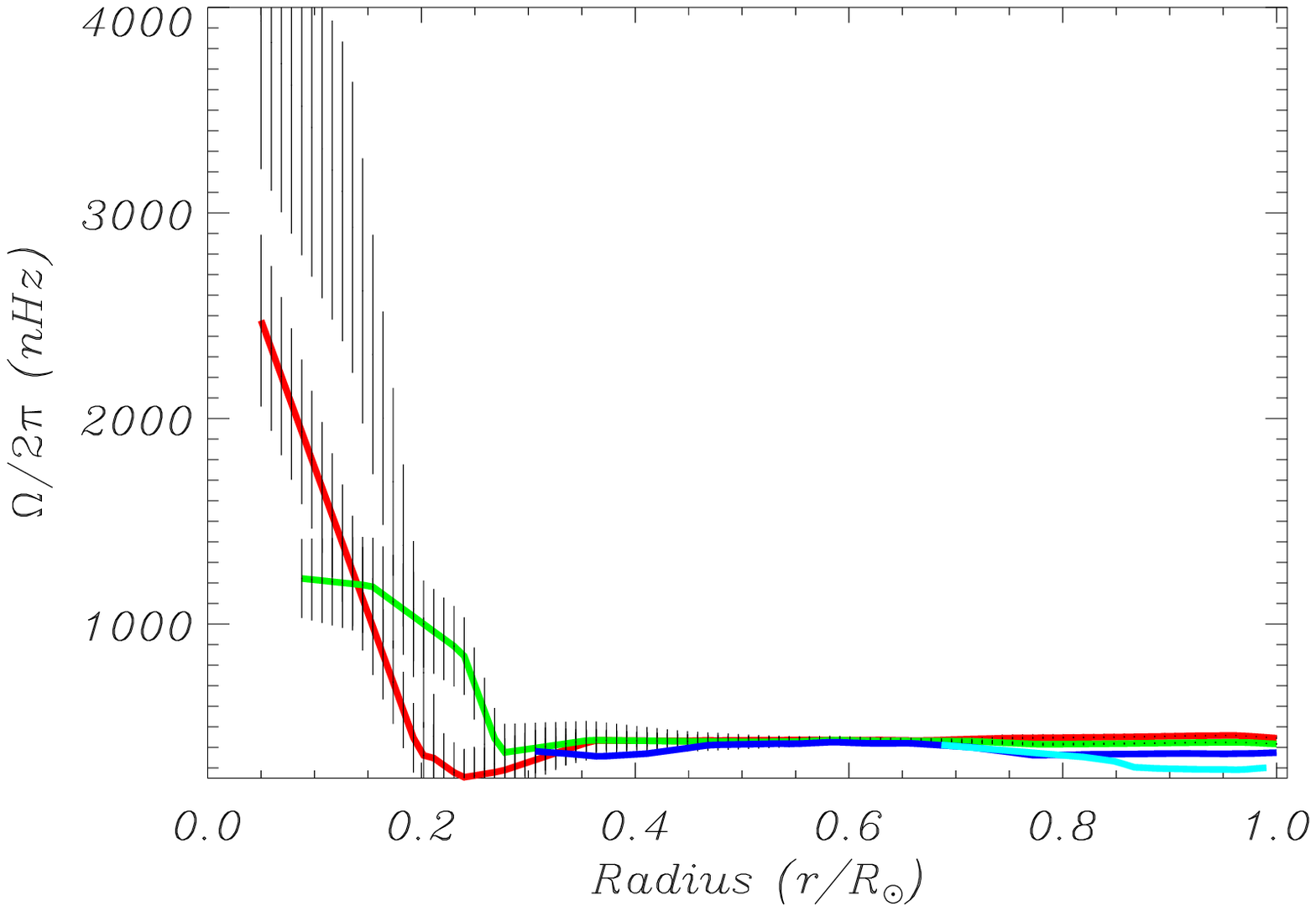}
\caption { \small { Left: Differences in squared sound speed and density between GOLF+MDI/SoHO and solar model predictions. Seismic model: full line + seismic error bars, SSM model ($ {- -}$) using the most recent updated physics. See \cite{Turck2011}. Right: Rotation profile extracted from the first gravity dipole modes and all the acoustic modes observed with the SoHO satellite. See \cite{Garcia2011}.}}
\normalsize
\label{fig:figure1}
\end{figure*}

Fig 1 left compares this profile to two models: the Solar Seismic Model  (SSeM), ajusted on the observed sound speed, and the SSM that integrates all the updated ingredients including the new photospheric CNO estimate, see the review of \cite{Turck2011b}  for more information (TCC2011). 

 From the same seismic data plus the first dipole gravity modes, we can also extract the rotation profile of Figure 1 right  down to the same region in the core. The error bars are still large due to the small number of modes detected. The latitudinal effect must be confirmed by the detection of more splittings which will improve such profile. Nevertheless one observes clearly \citep{Garcia2007,Garcia2011}  an increase in the solar core that succeeds to a quasi flat rotation in the rest of the radiative zone.

\subsection{A zoom on the solar core: neutrinos and gravity modes}
In the nuclear core,  three probes are now available: neutrinos emitted at different regions of the core, acoustic radial modes and gravity modes. We get now a lot of information from these 3 probes: 5 neutrino detectors sensitive to different energy of neutrinos, more than 20 radial acoustic modes and at least 6 gravity mode frequencies and the related splittings. 
In fact they all agree together through the predictions of the SSeM that has been built to reproduce only the sound speed on the whole radiative zone,  the information coming from one of the three probes: the acoustic mode frequencies. 

Figure 2 illustrates this point comparing the absolute sound speed with models and  the difference of the periods of the gravity modes between SSM and SSeM with the proximity of the observational values of GOLF. TCC2011 show also in their table 9 the excellent agreement between the prediction of the emitted neutrinos and the 5 neutrino detectors. 
The central temperature is strongly constrained by the boron neutrino flux and the central density by the gravity mode frequencies, so the  solar  values must be well reproduced by the seismic model values: T$_C$ = 15.75 $10^6$K  and $\rho_C$ =153.6 g/ cm$^3$. This information also useful for putting some constraints on the potential presence of dark matter \citep{Turck2011c}.

This seismic model is extremely useful for its predictions but it is not a physical model of Sun and solar-like stars as it is not able to reproduce the solar rotation profile (rotation effect is not included in the equations that describe SSeM or  SSM). It does not also predict any variability like the solar cycle or any other dynamical process that one needs to introduce in the radiative zone.
\begin{figure*}
\hspace{-1pc}
\includegraphics[width=18pc] {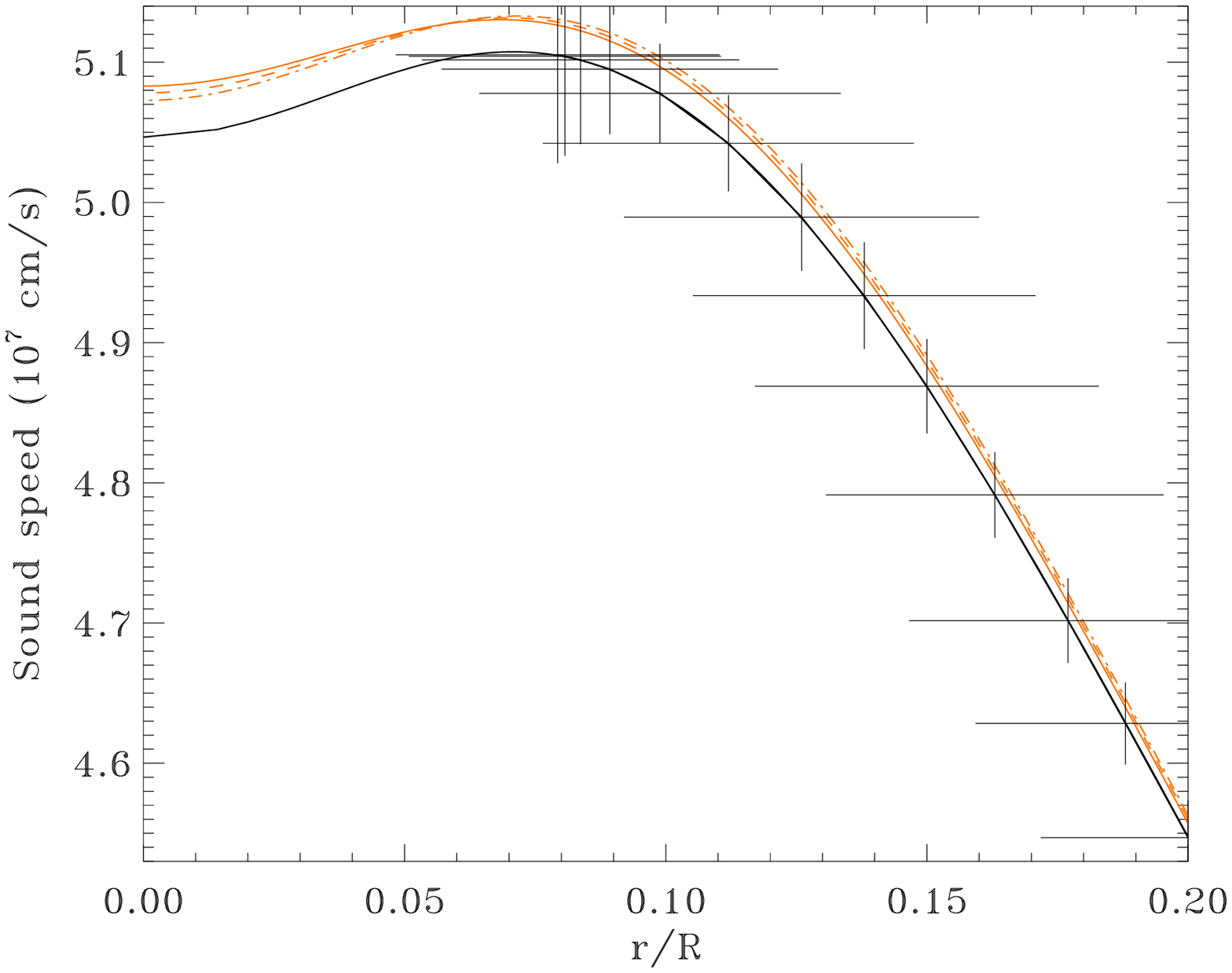}
\hspace{-1pc}
\includegraphics[width=15pc] {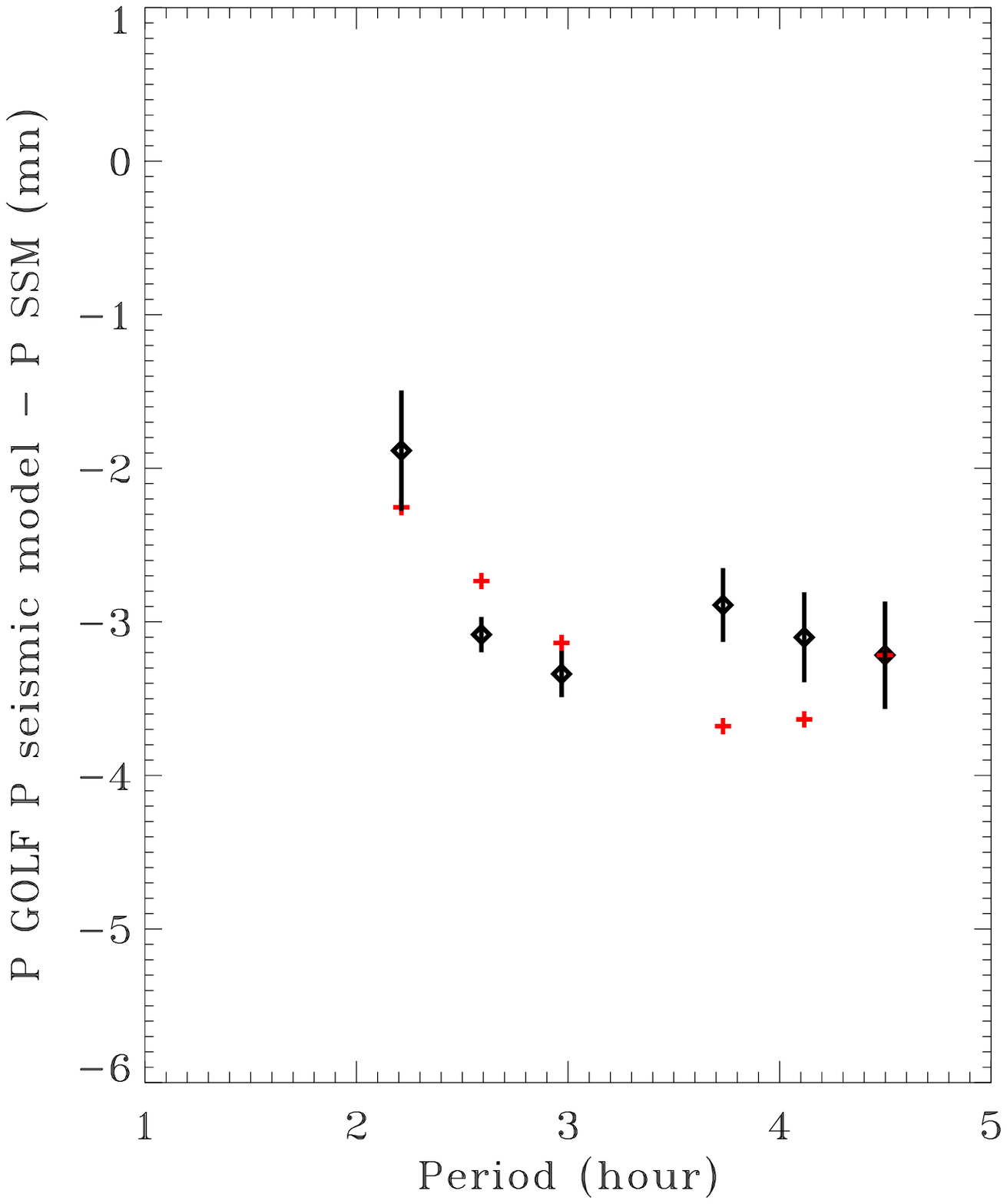}
\caption{ Left: Zoom on the absolute value of the sound speed in the core, the vertical error bars are multiplied by 100 for clarity. The extracted values are compared to the SSeM (in black) and to the SSM model in orange (dashed lines including also models with central luminosity increased by 2\%. See \cite{Turck2011}. Right: Difference in the dipole mode periods (expressed in minutes) between SSM and GOLF (diamonds with error bars) or SSeM (red crosses) and SSM. See also \cite{Turck2011c}.  }
\normalsize
\label{fig:figure2}
\end{figure*}
\section{The stars are rotating objects}
\begin{figure*}
\hspace{-2pc}
\includegraphics[width=18pc] {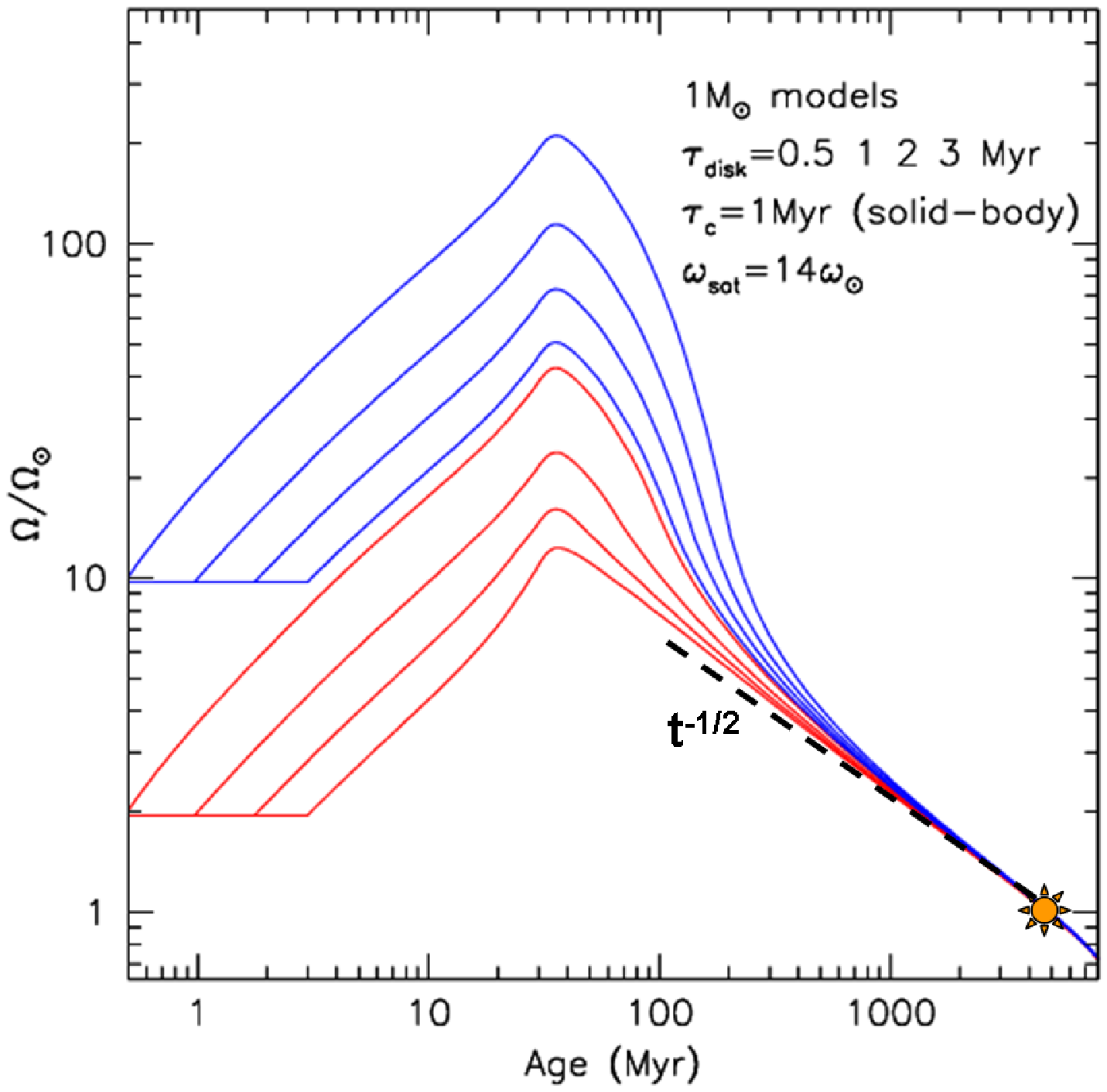}
\includegraphics[width=18pc] {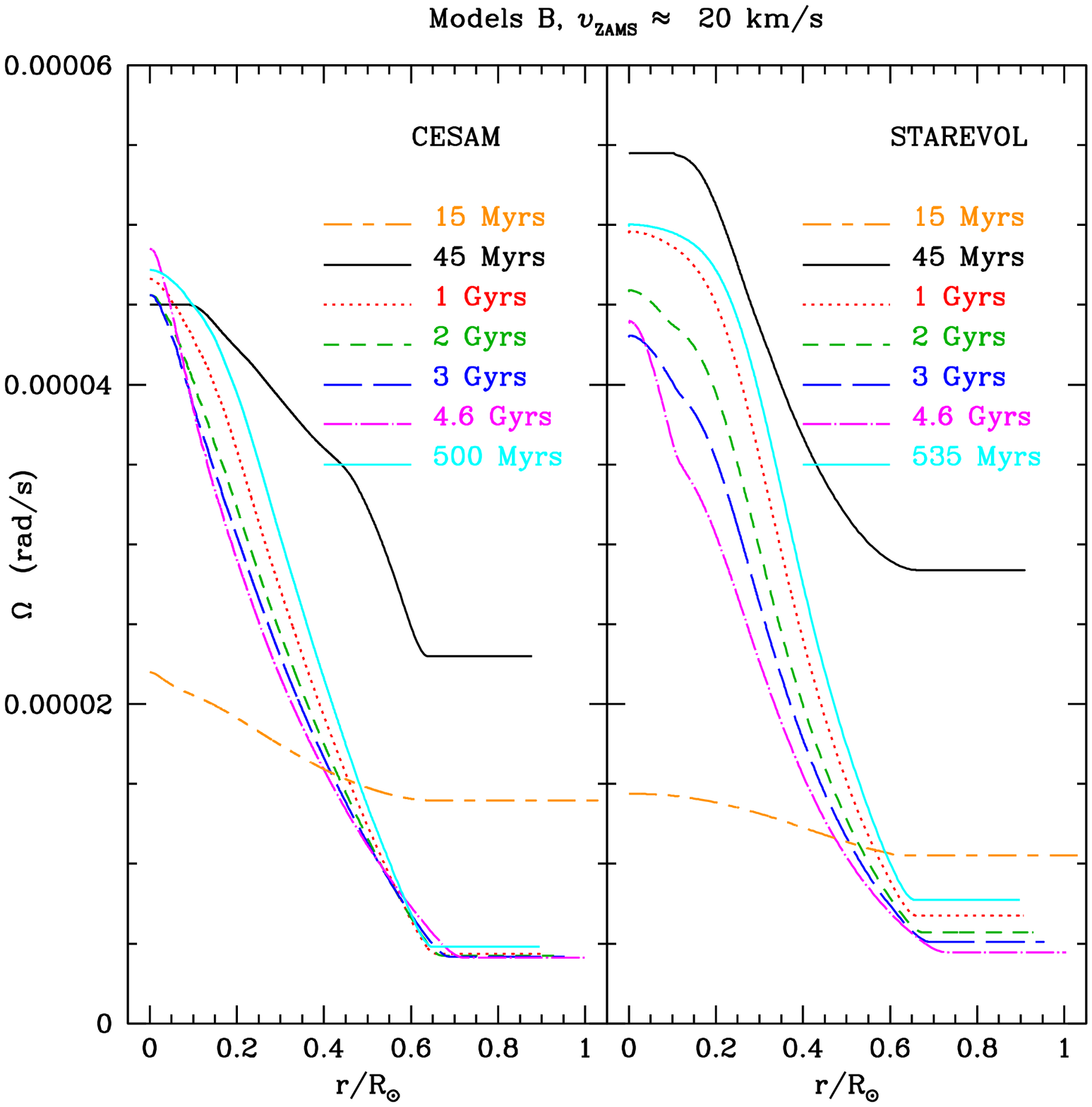}
\includegraphics[width=16pc] {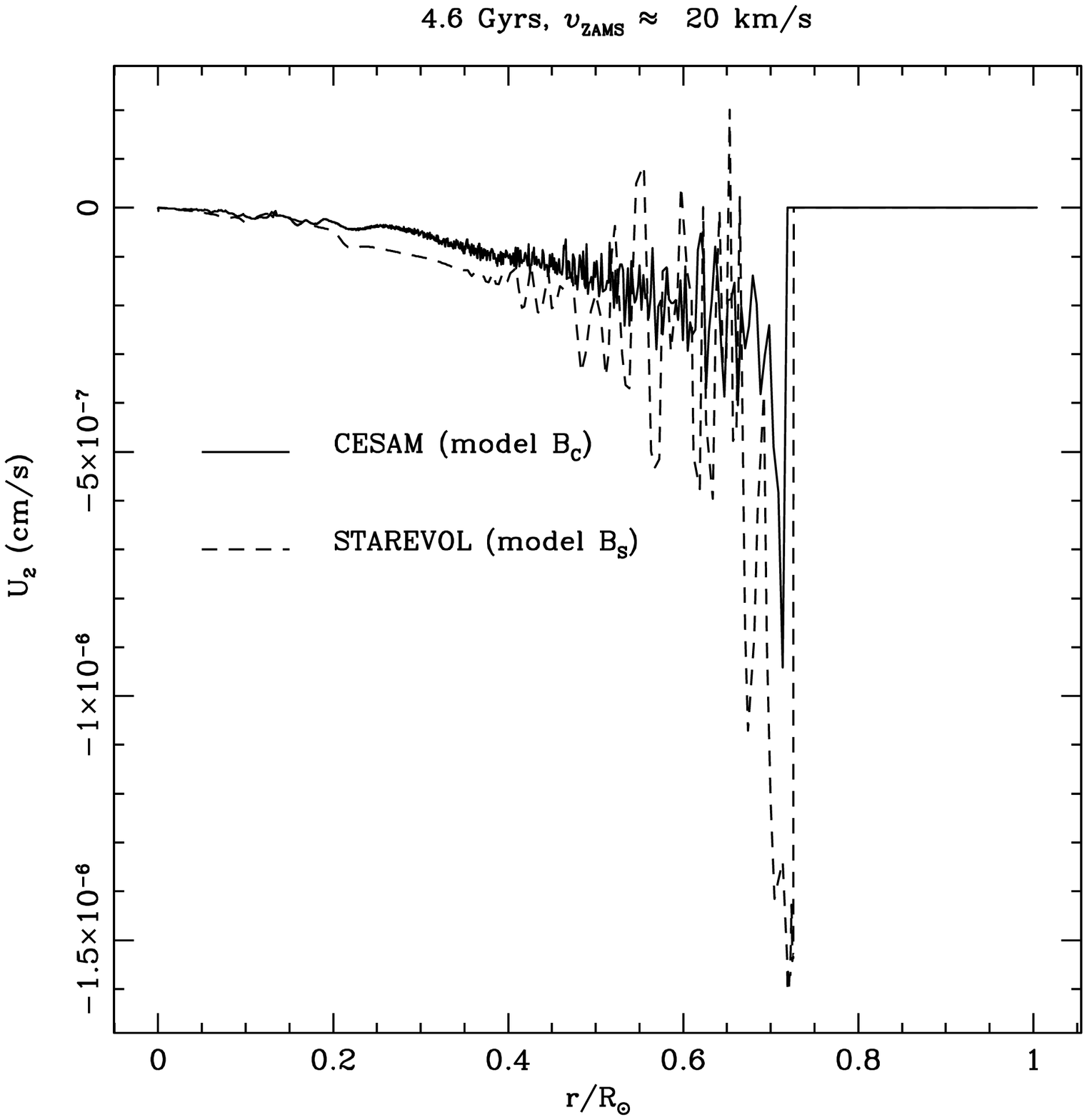}
\includegraphics[width=16pc] {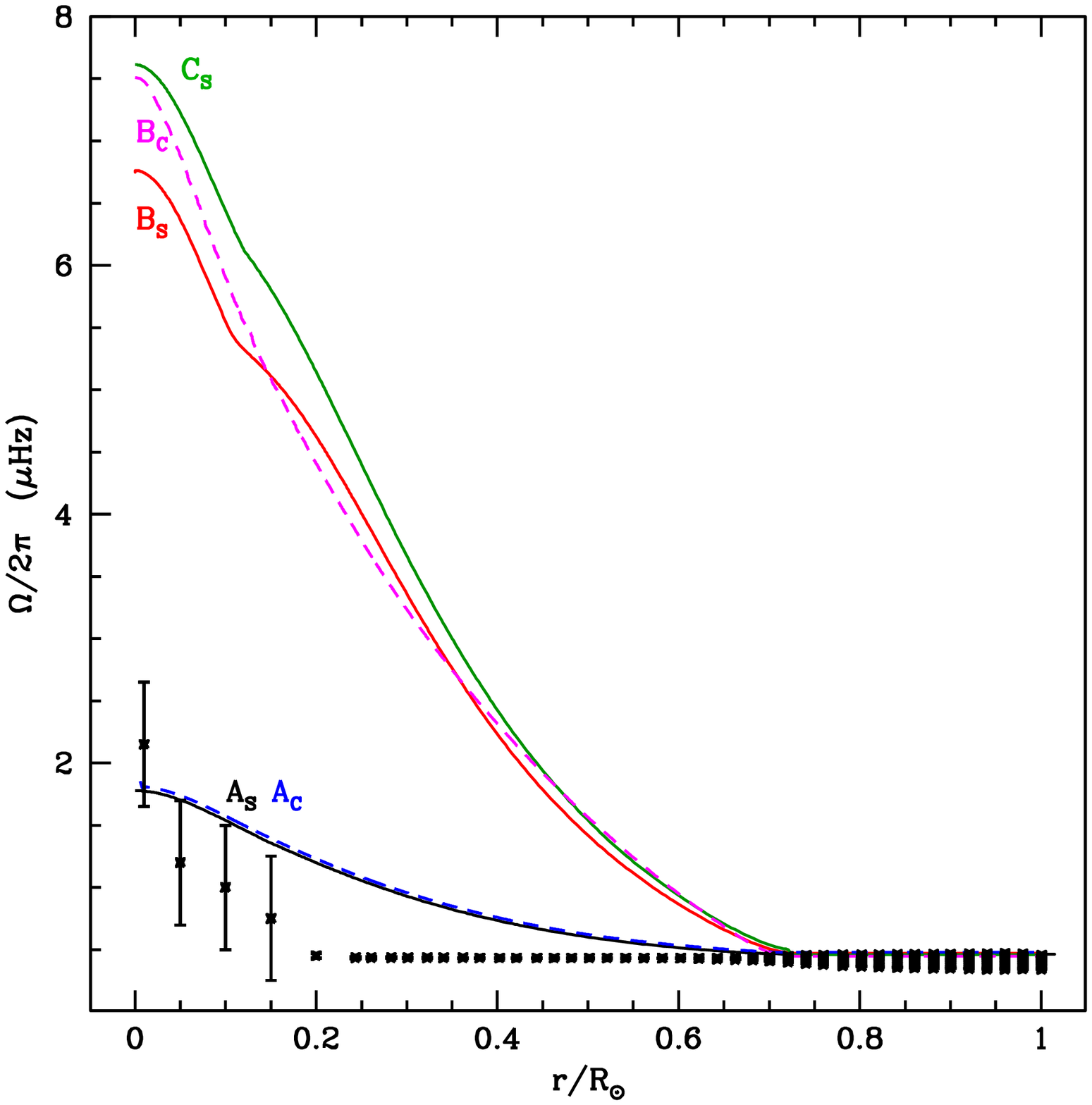}
\caption{ Top Left: Evolution of the rotation of young stellar objects of 1 M$_\odot$ in comparison with the solar present rotation. Two kinds of stars are considered: rapid and slow rotators, the Skumanich law is superimposed on the main sequence. From \cite{Bouvier}. Top Right: Time evolution of the internal rotation of a 1M$_\odot$ considered as a slow rotator reaching the MS at 20 km/s obtained by 2 different evolution codes: CESAM and STAREVOL the 45Myrs line corresponds to the arrival on the MS, then the star is slow down by magnetic braking applied at the surface. Bottom Left: Meridional circulation of the present Sun for the same model. Bottom Right: Comparison of the GOLF result to the different models (A, B and C) see text. From \cite{Turck2010a}.} 
\normalsize
\label{fig:figure2}
\end{figure*}
Rotation is the first hydrodynamical process  that exists in stellar radiative zones. It is generally ignored due to its small direct effect on the stellar structure except for massive stars.  For the first time,   we get  now the solar observed profile, so one needs to better explore the transport of momentum due to rotation along stellar  evolution. By chance, we begin also to have some hints from other stars thanks to the splitting of their mixed modes  for subgiant or red giant stars \citep{Deheuvels,Beck}. 

Let compare the obtained rotation profile to  models including rotation and transport of rotation \citep{Turck2010a}. We have built several models to follow the time  evolution of the rotation profile (see also in our work references of previous works). In these computations, we have followed the internal structure of the premain sequence evolution corresponding to  the decoupling of the star from the disk (see Figure 5).  We compare also our models to the observations  of \cite{Bouvier}. We have built two classes of models respectively called A and B or C. The first class (A) consists in  purely academic models starting with a small rotation  that evolves naturally toward the present external velocity of 2 km/s. The second class (B and C) reaches surface velocity of about 20 to 50 km/s  before the arrival on the main sequence and then are spin down by magnetic braking following the \cite{Skumanich} law down to the present solar rotation. Models B and C are compatible with  observations of \cite{Bouvier} (Figure 3 top left). Our  transport of momentum follows the prescription of Zahn (1992). Figure 3 illustrates our results and conclusions. One notices (Figure 3 top right) that the radial gradient of rotation is mainly established during the contraction phase,  the first 45 Myrs. Then the evolution of this profile is quite slow and slightly different from one code (for example CESAM) to another (STAREVOL). The two codes use the same transport of momentum prescription  but  the meridional circulation associated to the rotation transport becomes quickly extremely slow in comparison with what one can observe in the convective zone (Figure 3 bottom left). This fact contributes  to the difficulty to model such effect in a conventional evolutionary code and justifies slight numerical differences between them. 

Figure 3 bottom right summarizes the final results for the rotation internal profile. One can notice that the gradient observed for models B (20 km/s at the arrival on the main sequence and C (50 km/s) do not agree with the rotation profile observed by seismology. They show important gradient along the whole radiative zone. Models A agree better in the core but  models A do not respect the observations of young stars nor the spin down observed along the main sequence. Moreover in the two kinds of modeling, one never reaches a flat rotation profile outside the nuclear core as it is well established now by MDI results  \citep{Korzennik}.

So the progress done on the solar internal rotation profile by the first detection of gravity modes in the solar core and by acoustic modes in the rest of the radiative zone is presently not understood.  This year, asteroseismology adds new results that confirm the solar one: a gradient of about a factor 5 between core and surface in a subgiant  star \citep{Deheuvels}  and gradients of about a factor 10 in giants \citep{Beck}. These complementary observations  are compatible with a slow transport effect along the main sequence coupled to an accelerating of the core rotation in red giants in comparison to the surface rotation.
Such important results encourage certainly to go further on the dynamics of the radiative zone.

\newpage
\section{What we learn from young solar analogs}
\begin{figure}
\includegraphics[width=13pc] {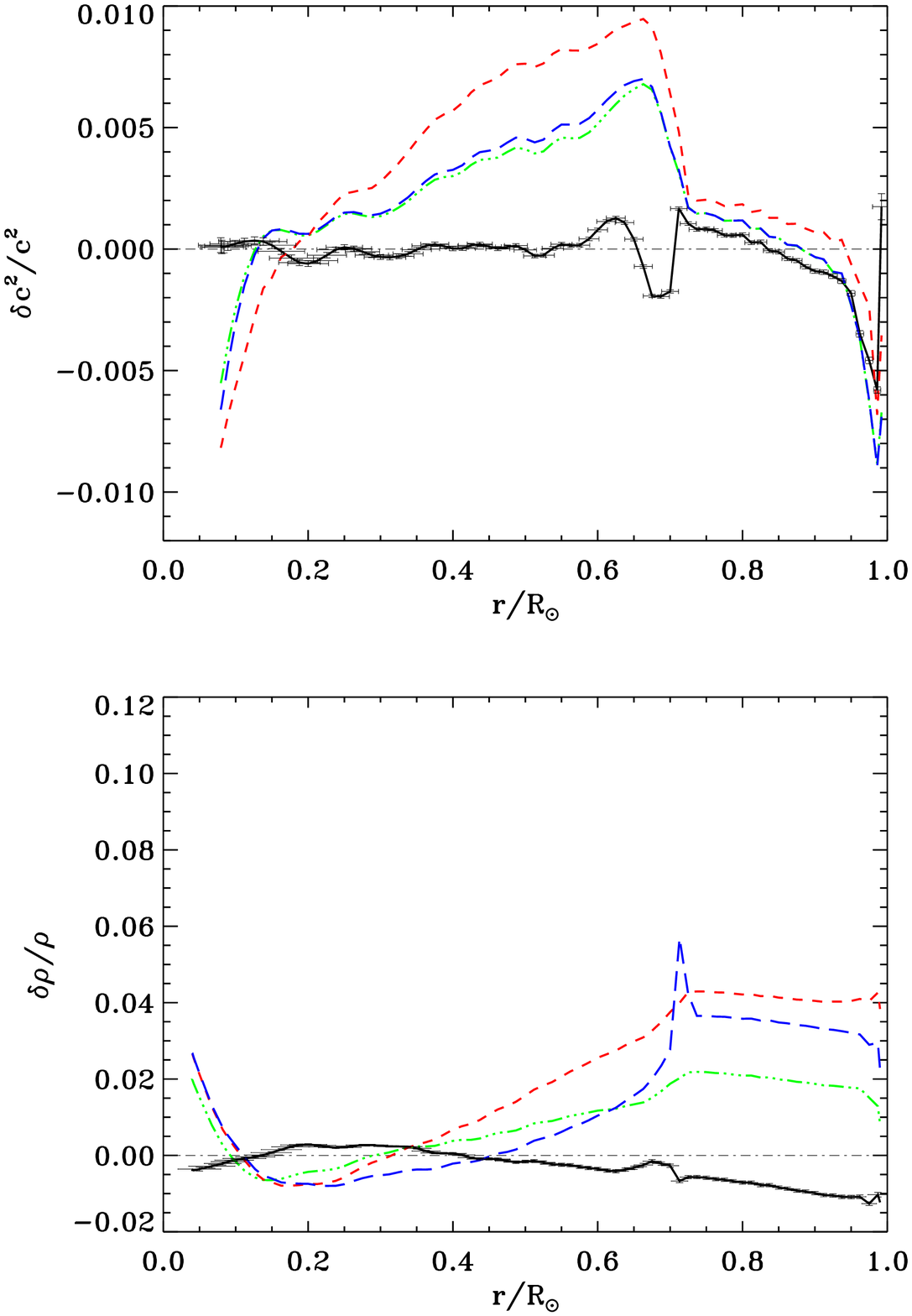} \hspace{2pc}
\includegraphics[width=13pc] {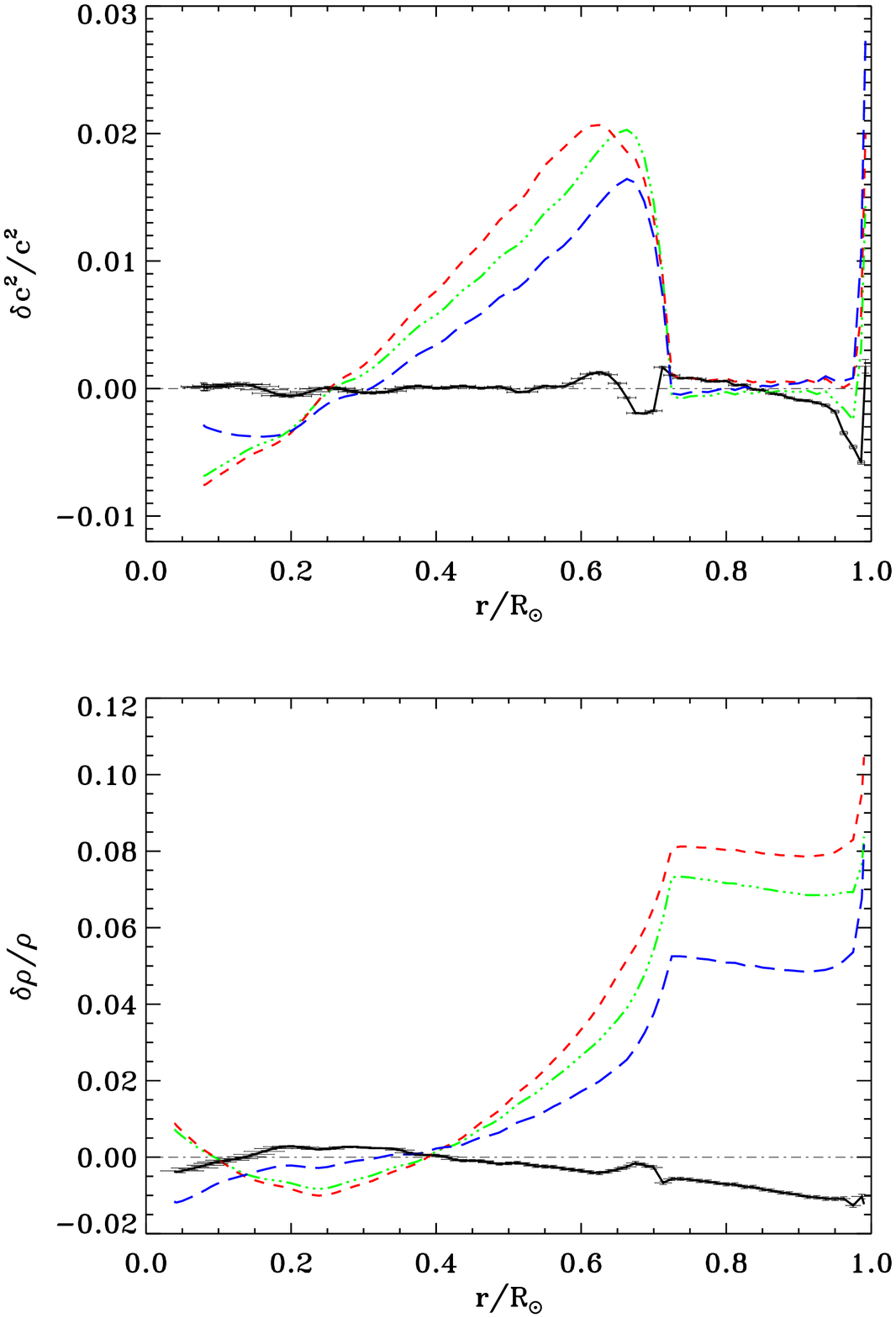}
\caption{  \small {Squared sound-speed and density differences 
between the helioseismic inversion of the acoustic mode frequencies obtained with the GOLF+MDI/SOHO instruments, and various solar models.
Left: The SSeM appears in black with error bars, the SSM models have been calculated for the GN composition (large dashed model and $- ...- $ models.) The small dashed curve corresponds to the SSM model modified by rotation and transport of momentum by rotation. From \cite{Turck2010a}. 
Right: The updated standard solar model including Asplund et al (2009)
is in green double dashed line, the SSM model with mixing in the tachocline is in red dashed line, the 
seismic solar model is in black line with the seismic error bars, and a model with mass loss in the early evolutionary phases following observations from young solar analogs is in blue long dashed line.  See \cite{Turck2011}.}}
\normalsize
\label{fig:figure4}
\end{figure}

The limitation of SSM appears  also in its lack to model the young solar analogs.
It is now well established that these stars are extremely active in their first stage and that their UV manifestation can be about 1000 times greater than the present solar activity. 

So if the internal rotation profile is largely influenced by that stage of evolution, it is important also to introduce the processes that are connected to that early activity. The energetic balance,  in the SSM framework,  is dominated by the nuclear production in the core and the transfer of energy by radiation in the whole radiative zone including the core. In real stars it is not clear that it is the same. The transformation of energy through kinetic energy, magnetic one  is not excluded during the early stage where the Sun was totally or mainly convective. Moreover the comparison between the central conditions of the Sun, verified by neutrinos and gravity modes, and the external luminosity let place for small difference of several $\%$ which cannot be understood inside the assumptions of the SSM where radiative transfer is instantaneous. See \cite{Turck2011}. 

Using young stellar analogous of the Sun and their time activity evolution laws, we deduce a greater initial mass than the present Sun and consequently an increase initial luminosity with important consequences for the formation of planets and for the induced present sound speed profile. 
Figure 4 shows the two effects mentioned in sections 3 and 4:  the introduction of the transport of momentum due to the rotation evolution that slightly increases the difference between SSM models and helioseismology. Here the comparison is done on the SSM that includes the old composition of Grevesse \& Noels (1993), it was not yet containing the updated of the CNO composition, in that case, the difference between SSM and SSeM is slightly reduced. On the right of the figure is shown the effect of introducing a stronger initial mass to simulate the initial activity and its associated mass loss. In this case the discrepancy between SSM and SSeM is slightly reduced. See details in \cite{Turck2011}. 
\begin{figure*}
\hspace{6pc}
\includegraphics[width=18pc] {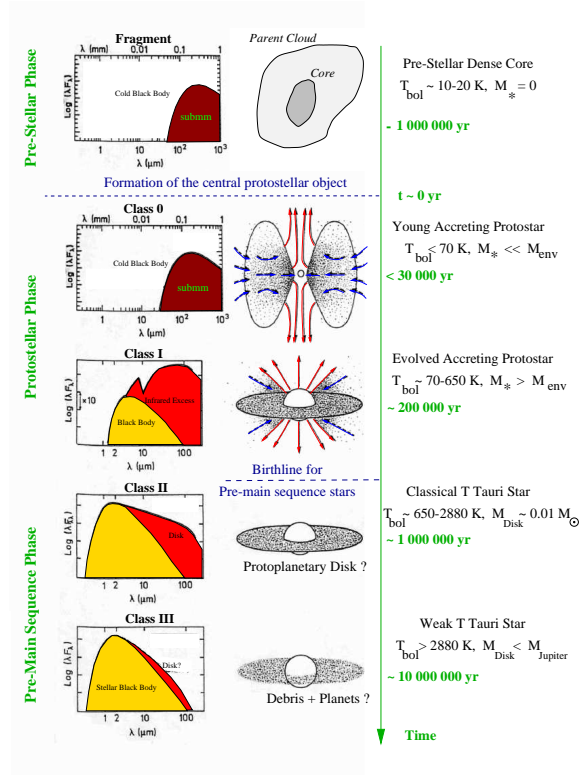}
\caption{ Representation of the evolution of young stellar objects. From \cite{Andre2002}.} 
\normalsize
\label{fig:figure5}
\end{figure*}
\section{The missing processes and the difficulty to introduce them}
Figure 5 shows the different steps in the formation of stars \citep{Andre2002} . If in phase 0, accretion dominates, in phase II corresponding to the PMS, solar analogous of 2 Myrs show that mass loss dominates accretion by an order of magnitude with complex magnetic configuration \citep{Donati2,Donati}.The building of real dynamical models of solar-like stars consists to introduce other processes like the presence of a fossil magnetic field built in the first stage during this extremely active phase when the star was still connected to the disk and largely convective. More and more indicators call for such a complex view of the radiative zone. We have begun to build some key ingredient to explore such possibility in using a magnetic field topology which takes into account poloidal and toroidal mixed fields \citep{Duez2010}. The difficulty is to estimate the strength of such a field and to find some indirect manifestation of this field: one could think to quadrupolar moments but they seem to be rather properly explained by the rotation profile \citep{Duez2011} or to some effects on the quadrupolar gravity modes but their identification is still puzzling \citep{Turck2004}. We shall begin with toy models of transport of momentum by magnetic field and diffusion as suggested by \cite{Mathis}.  Other stars, observed in seismology, will help in this new progress to build more realistic representation of solar-like stars and the bridge between young stellar object and main sequence stars.
\acknowledgements{This work was largely  supported by the CEA  and the space  CNES agency in their contribution to the GOLF instrument and through dedicated doctoral and postdoctoral grants. We are extremely grateful to ESA for their support on the SoHO satellite}.

\bibliographystyle{aspauthor}

\end{document}